\documentstyle[12pt]{article} 

\def\bm#1{\mbox{\boldmath $#1$}}
\def\ZR{{\bm{Z}}}
\def\CF{{\bm{C}}}
\def\Lpl{L_{\rm pl}}
\def\bop{\mathchoice{{\scriptstyle\circ}}{{\scriptstyle\circ}}{{\scriptscriptstyle\circ}}{\circ}}

\begin{document}

\begin{center}{\Large\bf Summary report of the workshop D2:\protect\\ 
Conceptual Issues, Foundational Questions \protect\\
and Quantum Cosmology}\footnote{To be published in the Proceedings of
the 15th General Relativity and Gravity Meeting.}
\end{center}

\begin{center}\large
{Hideo Kodama}
\end{center}

\begin{center}\large\sl
{Yukawa Institute for Theoretical Physics\protect\\
Kyoto University, Kyoto 606-8502, Japan}
\end{center}

\section{Introduction}

In this workshop rather diverging problems related to quantum
gravity and quantum cosmology were discussed. Since it is almost
impossible to summarize all these discussions in coherent and
integrated expressions, in this report, I give a brief description of
the content of each paper in order, with some additional comments on
their backgrounds. 
 
In order to give a readable presentation, I classified the oral and
poster papers presented in the workshop into six groups by their
contents.  The next section treats discussion on spacetime topology
and includes a single paper by S. Carlip. Section 3 includes three
papers on quantum cosmology by L.O. Pimentel and C. Mora, by
R. Mansouri and F. Naseri, and by the author(H.K.). Section 4 treats
papers on the interpretation problem in quantum gravity and cosmology
by D.P. Datta, by M. Kenmoku and others, by M. Morikawa and others, by
J. Larsson, and by A. Corichi and M.P. Ryan. Section 5 summarizes
discussions related to supersymmetry by R. Ferraro and D.M. Sforza, by
A.Yu. Kamenshchik and S.L. Lyakhovich, by H. Luckock, and by
H. Luckock and H. Farajollahi. Section 6 includes papers on various
non-standard approaches to quantum gravity and other arguments on
foundational problems presented by O. Richter, by P. Kuusk and others,
by J. Geddes, by R.S. Tung, by D.R. Finkelstein, by R.R. Karnik, and
by E.G. Mychelkin. The final section is devoted to the single paper by
S. Brave and T.P. Singh on the relation between the cosmic censorship
and the second law of thermodynamics.

\section{Spacetime Topology}


{\bf S. Carlip} reported on his very interesting observations on
topological fluctuations of spacetime in Euclidean quantum
gravity\cite{Carlip.S1997}.

He applied the saddle point approximation to the Euclidean
path-integral expression for the partition function of the pure
gravity system with cosmological constant $\Lambda(\not=0)$.  After
rescaling the metric so that each saddle point is represented by an
Einstein space whose Ricci scalar curvature is equal to $\pm12$, he
wrote the corresponding expression for $Z$ as a sum over the volume
$\tilde v$ of the rescaled space: $Z=\sum_{\tilde v} \rho({\tilde v})
\exp(9\tilde v/8\pi\Lambda\Lpl)$, where $\rho(\tilde v)$ is the number
of normalized Einstein spaces with volume $\tilde v$. One important
point here is that $\tilde v$ is a good index to classify topology
because each manifold seems to admit only a limited number of Einstein
metrics. Thus the sum over $\tilde v$ can be regarded as the sum over
topologies.

On the basis of this expression and estimations of $\rho(\tilde
v)$, he argued that quantum fluctuations of topology have quite
different nature for $\Lambda>0$ and $\Lambda<0$. In particular, he
showed that for $\Lambda<0$, $\ln\rho$ grows faster than $\tilde
v\ln\tilde v$. This implies that complicated topologies dominate
quantum fluctuations in this case. He further argued that
the Hartle-Hawking wavefunction may be sharply peaked at some special
3-topologies. 

Though Carlip's argument is still at the level of speculation, it
is quite fascinating and seems to have far-reaching implications.

\section{Quantum Cosmology}

Two papers discussed special quantum cosmological models in the
Wheeler-DeWitt approach.
 

First, {\bf L.O. Pimentel} and {\bf C\'esar Mora} considered a
spatially homogeneous scalar-tensor theory on the FRW
universe\cite{Pimentel.L&Mora1998}. They exactly solved the
Wheeler-DeWitt equation for this model and found that some special
subset of the solutions satisfy the boundary condition proposed by the
Hawking and Page, which requires that wave functions are regular at
degenerate geometries and fall off as the volume of 3-space
increases\cite{Hawking.S&Page1990}. They interpreted that these
solutions represent wormholes, although I could not understand the
reason.


Second, {\bf R. Mansouri} and {\bf F. Naseri} applied the
Wheeler-DeWitt quantization to a peculiar model proposed by
R. Mansouri and his
collaborators\cite{Khorrami.M&Mansouri&Mohazzab1996}, in which the
universe has a fractal structure and its dimension is dynamical, and
discussed the dependence of operator-ordering parameters on the
spatial dimension.


The author({\bf H.K.}) discussed the quantum dynamics of a spatially
compact Bianchi I model from a different viewpoint.  As is well know,
the canonical approach to quantum gravity based on the Wheeler-DeWitt
equation has various difficulties in mathematical formulation and
interpretation.  In order to eliminate these difficulties of the
conventional Dirac quantization, the author recently proposed a new
mathematically rigorous formalism, called the Web formalism, for
quantum dynamics of totally constrained systems\cite{Kodama.H1995}. In
this workshop, he talked about the relation of quantum dynamics in
this formalism to that in the Wheeler-DeWitt approach for the special
compact Bianchi model.

He first pointed out that in locally homogeneous pure gravity systems
obtained by compactifying Bianchi I models, their moduli freedom is
essentially frozen and their dynamics is described by a hamiltonian
system corresponding to the diagonal Bianchi I
model\cite{Kodama.H1997}. Then by restricting consideration to the
diagonal model on the basis of this result, he showed that quantum
dynamics in the Web formalism can be described by solutions to the
Wheeler-DeWitt equation only for special choices of the lapse
function. Further he discussed a general criterion for good time
variables or instant operators for this model.

\section{Interpretation Problem}

Finding natural correspondences between quantum theory and classical
theory plays an important role in constructing and interpreting
quantum theory. In particular, the extraction of time concept is the
major problem in the interpretation of dynamics of closed systems
including quantum gravity systems. In this workshop, three different
approaches to this problem were discussed.

The first one is the famous WKB approach, which corresponds to the
Born-Oppenheimer approximation used in the analysis of quantum
behavior of composite systems.  In this approach, one decomposes the
whole system ${\cal U}$ into a semi-classical part ${\cal C}$ and a
quantum part ${\cal Q}$ and writes the wave function $\Psi$ for the
whole system as the product of those for each part as $\psi\chi$. If
one applies the WKB approximation to the semi-classical part
$\psi=e^{iS}$, one obtains classical Hamiltonian-Jacobi trajectories,
on each of which a natural classical time can be defined, and the
evolution equation for $\chi$ is written as the Schr\"odinger equation
with respect to this time variable.


In this prescription, however, there is an ambiguity or a gauge
freedom in the choice of phase in the decomposition, which affects
the decomposed dynamics\cite{Datta.D1997}.  {\bf D.P. Data}
investigated the relation of this gauge freedom and quantum
fluctuations of the quantum subsystem and argued that under an
appropriate gauge fixing, one can pick up a natural time variable
which is related to quantum fluctuations of the system.


The second approach to the classical-quantum correspondence problem is
the de Broglie-Bhom interpretation for wave functions. In this
approach, one does not decompose the whole system ${\cal U}$ unlike in the
WKB approach.  Instead, one writes its wave function $\Psi$ simply
as $\Psi=A\exp(iS)$ and considers classical trajectories (dBB
trajectories) corresponding to the Hamilton-Jacobi function $S$. This
leads to equations of motion which are different from the classical
ones by a correction to the potential, called the quantum potential.
This quantum potential represents the effect of quantum fluctuations,
and if it is small for some variables describing a subsystem, this
subsystem behaves classically. The amplitude $A$ of $\Psi$ is
interpreted as giving the probability of each classical trajectory.

Mathematically speaking, this approach is exactly equivalent to the
standard formalism for quantum mechanics, because the wave function
$\Psi$ is assumed to satisfy the exact Schr\"odinger
equation. Physically speaking, however, it introduces a new
interpretation which is different from the conventional Copenhagen
interpretation: the dBB trajectories are regarded as objective
physical reality.

{\bf M. Kenmoku}, {\bf H. Kubotani}, {\bf E. Takasugi} and {\bf
Y. Yamazaki} reported their work on the application of this de
Broglie-Bohm interpretation to a spherically symmetric quantum black
hole\cite{Kenmoku.M&&1998}. They first solved the Wheeler-DeWitt
equation for this system and found generic exact solutions which are
eigenfunctions of the mass operator. Then they examined the behavior
of the dBB trajectories for these solutions and showed that the global
event horizon and the apparent horizon exactly coincide for a special
set of solutions whose dBB trajectories correspond to the classical
solution. They further calculated the quantum potential for these
special solutions and its influence on light rays propagating on the
spacetimes corresponding to the dBB trajectories. They also discussed
the dependence of the results on the operator ordering in the
Hamiltonian and the mass operators.


{\bf M. Morikawa}, {\bf F. Shibata}, {\bf T. Shiromizu} and {\bf
M. Yamaguchi} reported their recent work on another application of
this de Broglie-Bhom interpretation, which they called the potential
formalism.  Quantum tunneling of the universe occurs both in the
quantum creation of universe and in GUT phase transitions. This
quantum tunneling process is usually analyzed in the imaginary-time
formalism using the instanton method. However, this imaginary-time
formalism in general leads to a non-unitary evolution of the wave
function. M. Morikawa and his collaborators analyzed this quantum
tunneling process by the potential formalism and showed that this
formalism leads to a completely unitary description of the tunneling
process.


The third approach is the geometrical formulation proposed by
T.W.B. Kibble\cite{Kibble.T1979} and extended by A.~Ashtekar and
T.A.~Schilling\cite{Ashtekar.A&Schilling1997}. In this formulation,
utilizing the fact that the inner product of a Hilbert space ${\cal
H}$ induces a K\"ahler metric on the projective ray space of ${\cal
H}$, quantum theory is translated into a classical theory on this
K\"ahler space. Hence the classical phase is infinite-dimensional and
does not correspond to the phase space of the classical limit of
quantum theory. Apart from this peculiar feature, this formulation is
quite fascinating in that all concepts and physical quantities, such
as observables, spectrum, quantum fluctuations and probability, are
expressed in terms of a geometric language, and that the
correspondence between the quantum theory and the classical theory is
natural and exact.

{\bf A. Corichi} and {\bf M.P. Ryan} reported their preliminary work
on the application of this formulation to a minisuperspace quantum
gravity corresponding to the diagonal Bianchi I model with axial
symmetry.  In order to avoid technical difficulties arising from the
infinite-dimensionality of the classical phase space, they investigated
the reduction of the classical phase space to a finite-dimensional
subset consisting of coherent states.  They argued that the
investigation along this line would provide new insights on dynamics
and interpretation of quantum cosmology and gravity.


Finally, {\bf J.A. Larsson} discussed Bell's inequality in the hidden
variable theory in his poster paper, which is closely related to the
above interpretation problems. The main point of his work is to
generalize Bell's inequality by taking account of detector
inefficiency. In order to incorporate the detector inefficiency into
the formulation, he introduced the concept of `change of ensemble',
which provides both qualitative and quantitative information on the
nature of the `loophole' in the proof of the original Bell inequality. 
He showed that only models which contain change of ensemble lowers the
violation, and derived a bound on the violation, which does not depend
upon any symmetry assumptions such as constant efficiency, or the
assumption of independent errors.

\section{Supersymmetric Quantization}

In this section I summarize the papers which treated problems related
to supersymmetry.

 
Two papers discussed the BFV canonical quantization of the Einstein
gravity and related systems. First, {\bf R. Ferraro} and {\bf
D.M. Sforza} considered a constrained system with finite degrees of
freedom, which has one quadratic constraint $H_\perp$ in addition to
linear constraints, and discussed the operator-ordering problem in the
definition of the BRST charge operator $\Omega$. Utilizing the results
of their previous work\cite{Ferraro.R&Sforza1997}, they showed that
one can find a consistent ordering which gives $\Omega^2=0$, if the
potential term in $H_\perp$ is monotonically increasing along a
Killing vector of the supermetric.


{\bf A.Yu. Kamenshchik} and {\bf S.L. Lyakhovich} discussed a similar
problem for a specific but genuine Einstein gravity
system\cite{Kamenshchik.A&Lyakhovich1997}. They considered a
linearized Einstein gravity system with $N$ independent scalar fields
on locally Euclidean $d$-dimensional torus $T^d$. They first showed
that the constraint algebra is decomposed into a subalgebra
corresponding to area-preserving diffeomorphisms and a Virasoro-like
subalgebra, if the constraint operators are expanded by the Harmonic
tensors on $T^d$. Then, from the requirement $\Omega^2=0$, they
obtained the consistency condition, $N=30 + 5(d+1)(d-2)/2$. This
condition implies that the pure gravity system does not give a
consistent quantum theory in this framework. They also argued that
corrections introduced by going beyond the linear perturbation are of
order $\Lpl/V^{1/d}$, where $\Lpl$ is the Planck length and $V$ is the
space volume.


{\bf H. Luckock} discussed supersymmetry of quantum constraints in
locally supersymmetric theories. In general, local supersymmetry
cannot be represented on the configuration space of a theory, because
the transformation of configuration variables depends on momentum
variables. From this, they argued that a supergravity wave function
cannot be regarded as a superfield unlike that in globally
supersymmetric theories.


Finally {\bf H. Luckock} and {\bf H. Farajollahi} discussed the
stochastic quantization of locally supersymmetric theories and argued
that the Wick rotation of the physical time variable in such theories
plays the role of an extra fictitious time in the standard formulation
of stochastic quantization.

\section{Other Approaches to Quantum Gravity}

In addition to the investigations along major approaches to quantum
gravity and quantum cosmology presented so far, some unconventional
approaches were also discussed.


First, {\bf O. Richter} reported his preliminary work on quantum
gravity within the framework of non-commutative geometry of
A. Connes\cite{Connes.A1985}. He considered non-commutative geometry
on $M\times \Gamma_4$ with a discrete symmetry group $G$ acting on
$\Gamma_4$, where $M$ is an ordinary 4-manifold, and $\Gamma_4$ is a
set of four points. By formulating an analogue of the Einstein-Hilbert
action for this non-commutative geometry, he obtained discretized
versions of the $U(1)$ Kaluza-Klein model for $G=\ZR_4$ and the
non-linear $\sigma$ model for $G=T_4$(the tetrahedral group).


Second, {\bf P. Kuusk}, {\bf J. \"Ord} and {\bf E. Paal} discussed an
extension of the concept of translation to a generic curved
spacetime\cite{Kuusk.P&Ord1997}. The basic idea is to consider a
non-associative binary operation $x\cdot y$ for two points $x$ and $y$
in a Gaussian normal neighborhood at a point $e$ defined as $R_y
x\equiv x\cdot y=(\exp_y\bop \tau^e_y\bop \exp_e^{-1})(x)$, where
$\exp$ is the exponential mapping, and $\tau^e_y$ is the parallel
transport of vectors along the geodesic from $e$ to $y$. For a flat
spacetime, this binary operation coincides with the sum of the
position vectors for $x$ and $y$ with respect to the origin $e$. By
defining a momentum operator as the generator of the right translation
$R_a$, they constructed a non-standard Poisson algebra for the
position and the momentum of a particle in a curved spacetime and
discussed its quantization in the background spacetime with a weak
plane gravitational wave.






Other foundational problems related to quantum gravity were discussed
in poster papers. First, starting from an action which is expressed in
terms of a Dirac spinor 1-form (spin $3/2$-field) and a
$SL(2,\CF)$-connection and is equivalent to the Einstein-Hilbert 
action\cite{Tung.R&Jacobson1995},
{\bf R.S. Tung} defined a gravitational energy-momentum 3-form and
discussed its relevance to the issue of microscopic spacetime in
quantum gravity. Second, {\bf J. Geddes} proposed a new definition of the
functional integral for a special class of functionals of the form
$F[\phi]=\exp[\int dx f(\phi(x))]$. Finally, {\bf D.R. Finkelstein}
discussed a generalization of the general relativity principle, {\bf
R.R. Karnik} gave a speculation on spacetime properties, and {\bf E.G. 
Mychelkin} discussed some geometrodynamical aspects of spacetime.

\section{Cosmic Censorship}


The validity of the cosmic censorship hypothesis has a great
importance both in classical cosmology and astrophysics and in quantum
gravity. Since lots of counter examples are known at least in the
spherically symmetric case, there is a possibility that the cosmic
censorship hypothesis does not hold  in a pure mathematical sense
even for generic cases. 

From this observation, {\bf S. Barve} and {\bf T.P. Singh} argued that
the cosmic censorship may not be a consequence of the local laws of
physics and might be guaranteed by some additional general
principle. To be explicit, they proposed that the Weyl curvature
hypothesis proposed by R. Penrose\cite{Penrose.R1979} should be
imposed as a general principle to pick up a subset of initial
conditions which are actually realized in the universe. Here the term
`initial condition' is used in such a wider sense that the Weyl
curvature should vanish in the past limit along any past inextendible
causal curve. The motivation of their proposal is a close connection
between the Weyl curvature hypothesis and the second law of
thermodynamics. In order to see whether this proposal really works,
they calculated the behavior of the Weyl curvature near naked
singularities for known solutions and found that the Weyl curvature
diverges when one approaches the naked singularities along outgoing
null rays\cite{Barve.S&Singh1997}. This result implies that these
naked singularities do not appear in nature according to their
proposal.

\end{document}